\documentclass[preprint,amsmath,amssymb,superscriptaddress, floatfix,nofootinbib]{revtex4}
\usepackage{graphicx}
\usepackage{slashed}
\usepackage[autostyle]{csquotes}
\usepackage{xcolor}
\usepackage{soul}
\usepackage{hyperref}

\def\psla{p\kern-.45em/}
\def\ksla{k\kern-.45em/}
\def\qsla{q\kern-.45em/}
\def\delsla{\partial\kern-.45em/}
\def\nsla{n\kern-.45em/}

\def\ms{\ifmmode{\overline{\rm MS}} \else{$\overline{\rm MS}$} \fi}

\begin{document}
\begin{flushright}
KEK-TH-2621\quad\quad   TU-1231 
\end{flushright}

\title{One-loop QCD amplitudes in the Feynman-Diagram gauge }
\author{Kaoru Hagiwara}
\email{kaoru.hagiwara@kek.jp}
\address{KEK Theory Center and Sokendai, Tsukuba, Ibaraki 305-0801, Japan}
\author{Kentarou Mawatari }
\email{mawatari@iwate-u.ac.jp}
\address{Faculty of Education,  Iwate University, Morioka, Iwate 020-8550, Japan}
\address{Graduate School of Arts and Sciences, Graduate School of Science and Engineering, 
Iwate University, Morioka, Iwate 020-8550, Japan}
\author{Youichi Yamada}
\email{yoichi.yamada.c8@tohoku.ac.jp}
\address{Department of Physics, Tohoku University, Sendai 980-8578, Japan}
\author{Ya-Juan Zheng}
\email{yjzheng@iwate-u.ac.jp}
\address{Faculty of Education, Iwate University, Morioka, Iwate 020-8550, Japan}

\begin{abstract}
Scattering amplitudes for the massless QCD process, 
$q\bar{q}\to q^\prime\bar{q}^\prime$, are calculated 
in the one-loop order in the Feynman-Diagram (FD) gauge, 
where gluons are quantized on the light cone 
with opposite direction of the three-momenta. 
We find non-decoupling of the Faddeev-Popov ghosts 
and non-conventional UV singularities in dimensional regularization. 
The known QCD amplitudes with asymptotic freedom are 
reproduced only after summing propagator and vertex corrections. 
By quantizing gluons in the Feynman gauge on the FD gauge background, 
we obtain the one-loop improved FD gauge amplitudes.
\end{abstract} 
\maketitle

\section{Introduction}
\label{sec:intro}

Ref.\,\cite{Hagiwara:2020tbx} proposed a new form of the 
gauge boson propagator for massless gauge theories like QED and QCD, 
\begin{equation}
iG^{\rm FD}_{\mu\nu}(q) = \frac{i}{q^2+i0} \left( -g_{\mu\nu} 
+ \frac{q_{\mu}n_{\nu}(q)+n_{\mu}(q)q_{\nu}  }{n(q)\cdot q }  \right) , 
\label{fdpropagator1}
\end{equation}
where $n^{\mu}(q)$ is defined as 
\begin{equation}
n^{\mu}(q) = ({\rm sgn}(q^0), -q^i/|\vec{q}\,|) .
\label{FDlightcone}
\end{equation}
We use the notation 
$A^{\mu}=(A^0,\vec{A}\,)=(A^0,{A}^i)$ to separate 
time and space components of a four-vector. 
$n^{\mu}(q)$ is light cone, i.e. $n^{\mu}(q)n_{\mu}(q)=0$. 
Note that the propagator (\ref{fdpropagator1}) is not Lorentz covariant. 

Using the propagator (\ref{fdpropagator1}) for the photon and the gluon, 
it has been shown in ref.\,\cite{Hagiwara:2020tbx} that we can obtain 
helicity amplitudes which are free from subtle gauge cancellation 
among interfering Feynman diagrams.
This method was later extended \cite{Chen:2022gxv} to the electroweak theory, 
where massive gauge bosons are combined with associated Nambu-Goldstone modes 
forming 5-dimensional propagators. 
It has been found in refs.\,\cite{Hagiwara:2020tbx,Chen:2022gxv} that the absence of 
subtle cancellation among interfering Feynman diagrams and the collinear properties 
of individual diagram are common in the massless~\cite{Hagiwara:2020tbx} 
and in the massive~\cite{Chen:2022gxv} gauge theories. 
Because of these common properties,
\footnote{The propagator was called \textquoteleft parton shower gauge\textquoteright~ in 
ref.\,\cite{Hagiwara:2020tbx} , because the magnitude of 
individual Feynman diagram agrees with parton splitting 
amplitudes~\cite{Hagiwara:2009wt} in the collinear limit.
It was later renamed as Feynman-Diagram gauge in 
refs.\,\cite{Chen:2022gxv,Chen:2022xlg} because the term \textquoteleft parton 
shower gauge\textquoteright~was used 
in refs.\,\cite{Nagy:2007ty,Nagy:2014mqa} for a specific light-cone gauge.} 
eq.~(\ref{fdpropagator1}) is named
\textquoteleft Feynman-Diagram (FD) gauge' in ref.\,\cite{Chen:2022gxv}. 

It has later been shown in  ref.\,\cite{Chen:2022xlg} that 
the propagator (\ref{fdpropagator1}), as well as its generalization to massive 
gauge bosons~\cite{Chen:2022gxv}, can be derived from the gauge fixing term 
similar to that in the light-cone gauge \cite{Leibbrandt:1987qv}. 

In this paper, we study radiative corrections for massless gauge theories 
in the FD gauge. 
The rest of this paper is organized as follows. 
In section~\ref{sec:Feynmanrule}, we show the relevant Feynman rules in the FD gauge 
for loop calculation. 
Section \ref{sec:amplitude} gives details of the one-loop scattering amplitudes 
for a massless quark scattering process, $q\bar{q}\to q'\bar{q}'$, in the FD gauge. 
Section~\ref{sec:bkg} shows that by quantizing gluons in the Feynman gauge 
on the FD gauge background, we can obtain one-loop corrected FD gauge amplitudes. 
Section~\ref{sec:summary} summarizes our finding, and some technical details 
of the loop integrals are given in appendices \ref{sec:AppA} and \ref{sec:AppB}. 

\section{Feynman rules in the FD gauge }
\label{sec:Feynmanrule}

We work in QCD with massless quarks. The Lagrangian takes the form 
\begin{equation}
{\cal L} = -\frac{1}{4}F^a_{\mu\nu}F^{a\mu\nu} 
+ \sum_q i \bar{q}_i \gamma^{\mu}(\partial_{\mu}\delta_{ij} +ig A_{\mu}^a (T^a)_{ij}) q_j 
+{\cal L}_{GF}+{\cal L}_{FP} . 
\label{lagrangian} 
\end{equation}
In this section, we give the forms of the gauge fixing term ${\cal L}_{GF}$ and the 
Faddeev-Popov (FP) ghost term ${\cal L}_{FP}$ corresponding to 
the FD gauge propagator (\ref{fdpropagator1}). 

Following ref.\,\cite{Chen:2022xlg}, we consider a gauge fixing 
\begin{equation}
{\cal L}_{GF} = -\frac{1}{2\xi}  (F^a[A])^2 ,
\label{FDgaugefixing0a}
\end{equation}
with 
\begin{equation}
F^a[A] = \hat{n}^{\mu}(\partial) A_{\mu}^a , 
\label{FDgaugefixing0}
\end{equation}
and the gauge parameter $\xi$. 
Here $\hat{n}^{\mu}(\partial)$ is a differential operator that may be 
Lorentz non-covariant and even nonlocal, which was not manifestly written 
in ref.\,\cite{Chen:2022xlg}.  

The kinetic term for the gluon is 
\begin{equation}
{\cal L}_K= \frac{1}{2}A^{a\mu} \left( 
g_{\mu\nu} \partial^2 -\partial_{\mu}\partial_{\nu} 
-\frac{1}{\xi} \overleftarrow{\hat{n}}_{\mu}\hat{n}_{\nu} \right) A^{a\nu} ,
\end{equation}
with $\overleftarrow{\hat{n}}_{\mu}=-\hat{n}_{\mu}$. 
The equation of motion (EOM) of $A$, with the source term $J^a_{\mu}A^{a\mu}$ 
added, is then 
\begin{equation}
\left( 
g_{\mu\nu} \partial^2 -\partial_{\mu}\partial_{\nu} 
+\frac{1}{\xi} \hat{n}_{\mu}\hat{n}_{\nu} \right) A^{a\nu}= -J^a_{\mu} .
\label{fdeom}
\end{equation}

In moving to the momentum space, 
we set the momentum space representation of $\hat{n}^{\mu}(\partial)$ as $-in^{\mu}(q)$
and the EOM\,\eqref{fdeom} gives the gluon propagator,
\begin{equation}
i\delta^{ab}G_{\mu\nu}^{\rm FD}(q)|_{\xi} = 
\frac{i\delta^{ab}}{q^2+i0} \left( -g_{\mu\nu}+\frac{q_{\mu}n_{\nu}(q)+n_{\mu}(q)q_{\nu}} 
{n(q)\cdot q} -\frac{\xi q^2 q_{\mu} q_{\nu} }{(n(q)\cdot q)^2} 
\right) . 
\label{gaugepropagator-xi}
\end{equation}
Hereafter we set $\xi\to 0$ and obtain
\begin{equation}
i\delta^{ab}G_{\mu\nu}^{\rm FD}(q) = 
\frac{i\delta^{ab}}{q^2+i0} \left( -g_{\mu\nu}+\frac{q_{\mu}n_{\nu}(q)+n_{\mu}(q)q_{\nu}} 
{n(q)\cdot q}  
\right)  \equiv 
\frac{i\delta^{ab}}{q^2+i0}P_{\mu\nu}^{\rm FD}(q) . 
\label{gaugepropagator}
\end{equation}
Eq.\,(\ref{gaugepropagator}) gives the gluon propagator (\ref{fdpropagator1})
in the FD gauge. 
Note that (\ref{gaugepropagator}) explicitly breaks Lorentz invariance, 
while keeping space rotational invariance for the light-cone vector of~\eqref{FDlightcone}. 

To calculate loop corrections in the FD gauge, 
we also need to determine the Lagrangian for 
the FP ghosts $(c^a,\bar{c}^a)$ associated with the 
gauge fixing (\ref{FDgaugefixing0a}, \ref{FDgaugefixing0}). 
In the coordinate space, the Lagrangian for the FD ghosts is~\cite{Srednicki:2007qs}
\begin{align}
{\cal L}_{FP}  & = i \bar{c}^a \frac{\delta F^a[A] }{\delta A^b_{\mu}} (D_{\mu} c)^b 
\nonumber \\ 
& = i \bar{c}^a \hat{n}^{\mu}  (D_{\mu} c)^a 
\nonumber \\ 
& = i \bar{c}^a \hat{n}^{\mu}  \partial_{\mu} c^a 
- ig f^{abc} \bar{c}^a \hat{n}^{\mu} A^b_{\mu} c^c ,  
\end{align}
where $(D_{\mu}c)^a=\partial_{\mu}c^a-gf^{abc}A^b_{\mu}c^c$ is 
the covariant derivative of the ghost $c$. 
The propagator and $c\bar{c}A_{\mu}$ coupling of the FP ghosts are then 
given as 
\begin{equation}
\langle c^a(q)\bar{c}^b(-q) \rangle = 
i\delta^{ab} G_{FP}(q) = -\frac{\delta^{ab} }{n(q)\cdot q}
=-\frac{\delta^{ab} }{|q^0|+|\vec{q}\,| }, 
\label{FPFD1}
\end{equation}
\begin{equation}
i\Gamma (  \bar{c}^a(-p)A^{b\mu}(p-q)c^c(q) ) = -igf^{abc}n^{\mu}(p) ,
\label{FPFD2}
\end{equation}
respectively. 
Note that unlike in the light-cone gauge in which $n^\mu$ is common for all gluons, 
the FP ghosts don't decouple from the amplitudes.


\section{Four-quark scattering amplitudes in the FD gauge}
\label{sec:amplitude}

\begin{figure*}
  \center
\includegraphics[width=1\textwidth]{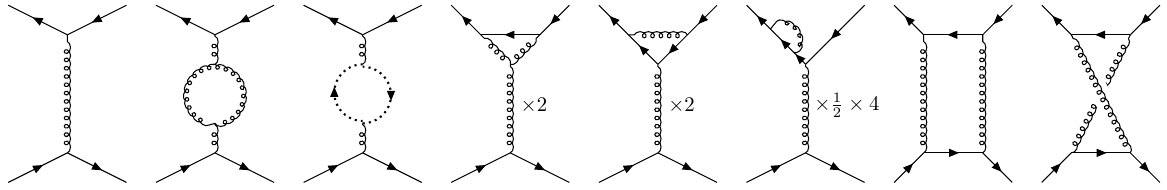}
(a)\hspace*{15mm}(b)\hspace*{15mm}(c)\hspace*{15mm}(d)\hspace*{15mm}
(e)\hspace*{15mm}(f)\hspace*{15mm}(g)\hspace*{15mm}(h)
\caption{Feynman diagrams contributing to $q\bar q\to q'\bar q'$ at the tree and 
the one-loop order.}
\label{fig:diagram_qqqq}
\end{figure*}

To discuss loop corrections in the FD gauge, we use 
the massless quark scattering $q\bar{q}\to q'\bar{q}'$ $(q\neq q')$ 
and calculate the amplitudes with one-loop corrections by gluons, 
as shown in Fig.~\ref{fig:diagram_qqqq}. 

The tree-level amplitudes Fig.\,1(a) of the process
\begin{eqnarray}
q_i(p_1)+\bar{q}_j(p_2)\to g^a(q)\to q'_l(p_3)+\bar{q}'_m(p_4),
\end{eqnarray} 
with color indices for quarks ($i,j,l,m$) and gluon ($a$), 
is 
\begin{equation}
i{\cal M}^{(a)} = -ig^2 \bar{v}(p_2)(T^a)_{ji}\gamma^{\mu}u(p_1)
\frac{ P^{\rm FD}_{\mu\nu}(q) }{ q^2 +i0}
\bar{u}(p_3)(T^a)_{lm}\gamma^{\nu}v(p_4) . \label{eq1}
\end{equation}
Here $iP^{\rm FD}_{\mu\nu}(q)/(q^2+i0)$ is the gluon propagator 
in the FD gauge (\ref{gaugepropagator}). 

Relations $\bar{v}(p_2)\qsla u(p_1)=\bar{u}(p_3)\qsla v(p_4)=0$ follow 
from the EOM of the quarks. 
As a consequence, the $n_{\mu}q_{\nu}+q_{\mu}n_{\nu}$ parts of $P^{\rm FD}_{\mu\nu}$ 
do not contribute 
to the amplitude (\ref{eq1}), as expected from 
the gauge independence of the 
on-shell scattering amplitudes.  

We now evaluate the one-loop corrections to the amplitude by gluons. 
Before showing explicit calculations, 
we review the structure of the loop corrections. 
As shown in Fig.~\ref{fig:diagram_qqqq}, they consist of the corrections by 
gluon self-energies (b, c), 
quark-quark-gluon vertex corrections (d, e) 
with wave function corrections of quarks (f), 
and four-quark box corrections (g, h). 

The gluon loop corrections to the amplitude (\ref{eq1}) are then 
\begin{equation}
i{\cal M}({\rm corr}) = i{\cal M}^{(b+c)} + i{\cal M}^{(d+e+f)} 
+ i{\cal M}^{(g+h)} .  \label{eq4}
\end{equation}
In terms of the gluon self energy 
$i\Pi^{\mu\nu}(q)=i\Pi^{(b)\mu\nu}(q)+i\Pi^{(c)\mu\nu}(q)$,  
$i{\cal M}^{(b+c)}$ is expressed  as 
\begin{equation}
i{\cal M}^{(b+c)} = 
 ig^2 \bar{v}(p_2)(T^a)_{ji}\gamma^{\mu}u(p_1)
\frac{P^{\rm FD}_{\mu\lambda}(q)}{q^2+i0} 
\Pi^{\lambda\rho}(q)
\frac{P^{\rm FD}_{\rho\nu}(q)}{q^2+i0} 
\bar{u}(p_3)(T^a)_{lm}\gamma^{\nu}v(p_4) . \label{eq5}
\end{equation}
Substituting the explicit form of $P^{\rm FD}_{\mu\nu}$ (\ref{gaugepropagator}) 
and EOMs for quarks, the correction is expressed as 
\begin{align}
i{\cal M}^{(b+c)} = 
& -ig^2 \bar{v}(p_2)(T^a)_{ji}\gamma^{\mu}u(p_1) 
\left[ 
-\frac{1}{q^4} \Pi_{\mu\nu}(q)
 \right. 
\nonumber \\ 
& + \frac{n_{\mu}(q)}{q^2 (n(q)\cdot q)}(q^{\lambda} \Pi_{\lambda\nu}(q)) \frac{1}{q^2} 
+ \frac{1}{q^2} ( \Pi_{\mu\rho}(q) q^{\rho}) \frac{n_{\nu}(q)}{q^2(n(q)\cdot q)} 
\nonumber \\ 
& \left. 
-\frac{ n_{\mu}(q) }{q^2(n(q)\cdot q)} 
(q^{\lambda} q^{\rho} \Pi_{\lambda\rho}(q) )
\frac{ n_{\nu}(q) }{q^2 (n(q)\cdot q)} \right] 
\bar{u}(p_3)(T^a)_{lm}\gamma^{\nu}v(p_4) . \label{eq7}
\end{align}
Furthermore, by using the explicit form of $n^{\mu}(q)$,  we have 
\begin{align}
i{\cal M}^{(b+c)}  =&
-ig^2 \bar{v}(p_2)(T^a)_{ji}\gamma^{\mu}u(p_1) \frac{1}{q^4}
\bar{u}(p_3)(T^a)_{lm}\gamma^{\nu}v(p_4) 
\nonumber \\
& \times \left[ -\Pi_{\mu\nu}(q)  +\frac{{\rm sgn}(q^0)}{|\vec{q}\,|}
(t_{\mu}q^{\rho}\Pi_{\rho\nu}(q) + \Pi_{\mu\sigma}(q) q^{\sigma}t_{\nu}) 
-t_{\mu}t_{\nu}\frac{1}{|\vec{q}\,|^2}q^{\rho}q^{\sigma}\Pi_{\rho\sigma}(q) \right] , 
\label{ampbc}
\end{align}
where $t^{\mu}=(1,0,0,0)$ is a constant vector. 
In eq.~(\ref{ampbc}) the following relation from the quark EOM, 
\begin{align}
 \frac{n_{\mu}(q)}{n(q)\cdot q}\bar{v}(p_2) \gamma^{\mu} u(p_1) 
&= 
\frac{1}{|q^0|+|\vec{q}\,|}\bar{v}(p_2) \left[ {\rm sgn}(q^0)\gamma^0 
+q^i\gamma^i\frac{1}{|\vec{q}\,|} \right] u(p_1)  
\nonumber \\
& =\frac{1}{|q^0|+|\vec{q}\,|}\bar{v}(p_2) \left[ 
{\rm sgn}(q^0)\gamma^0 +q^0\gamma^0\frac{1}{|\vec{q}\,|} \right]  u(p_1) 
\nonumber \\
&
=\frac{{\rm sgn}(q^0)}{|\vec{q}\,|}\bar{v}(p_2) \gamma^0 u(p_1) ,
\label{nslaid} 
\end{align} 
and a similar relation for $\bar{u}(p_3)\gamma^{\nu}v(p_4)$ are used. 
As we will see later, 
contrary to the case in the covariant gauges, 
$q^{\mu} \Pi_{\mu\nu}(q)$ in the FD gauge does not vanish in general. 

Similarly, $i{\cal M}^{(d+e+f)}$ is expressed in terms of the 
${q}qg$ vertex functions $i\Gamma^{\mu}$ and $i\Gamma^{\nu}$ , 
which is the sum of the 1PI vertex corrections (d+e) and 
quark wave function corrections (f), as 
\begin{align}
i{\cal M}^{(d+e+f)} = 
& \bar{v}(p_2)(T^a)_{ji}i\Gamma^{\mu}(-q,p_1,p_2)u(p_1)
\frac{iP^{\rm FD}_{\mu\nu}(q)}{q^2} 
(-ig)\bar{u}(p_3)(T^a)_{lm}\gamma^{\nu}v(p_4)
\nonumber \\
& + (-ig) \bar{v}(p_2)(T^a)_{ji}\gamma^{\mu}u(p_1)
\frac{iP^{\rm FD}_{\mu\nu}(q)}{q^2} 
\bar{u}(p_3)(T^a)_{lm}i\Gamma^{\nu}(q,-p_4,-p_3)v(p_4) . \label{eq6} 
\end{align}
By using (\ref{nslaid}) again, we have 
\begin{align}
i{\cal M}^{(d+e+f)} = & 
ig^2 \bar{v}(p_2)(T^a)_{ji}\Gamma^{\mu}(-q,p_1,p_2) u(p_1) \frac{1}{q^2}
\left( -g_{\mu\nu} + q_{\mu} \frac{{\rm sgn}(q^0)}{ |\vec{q}\,| }t_{\nu}  \right) 
\bar{u}(p_3)(T^a)_{lm} \gamma^{\nu} v(p_4) 
\nonumber \\
+ & 
ig^2 \bar{v}(p_2)(T^a)_{ji} \gamma^{\mu} u(p_1) \frac{1}{q^2}
\left( -g_{\mu\nu} + t_{\mu} \frac{{\rm sgn}(q^0)}{ |\vec{q}\,| } q_{\nu}  \right) 
 \bar{u}(p_3)(T^a)_{lm} \Gamma^{\nu}(q,-p_4,-p_3) v(p_4)  .
\label{mvertex3} 
\end{align}


\subsection{UV divergent parts of the corrections}

We now evaluate the UV divergence of each part of the 
gluon loop correction (\ref{eq4}) in the FD gauge. 

First, 
gluon self energy by gluon loop (b) and by FP ghost loop (c) are 
\begin{align}
i\delta^{ab}\Pi^{(b)}_{\mu\nu}(q) = & -\frac{1}{2}f^{acd}f^{bcd}g^2 
\int \frac{d^D k}{(2\pi)^D}
[(-k+q)_{\rho}g_{\mu\lambda}+(2k+q)_{\mu}g_{\lambda\rho}+(-2q-k)_{\lambda}g_{\mu\rho}]
\nonumber \\
& \times 
[(k-q)_{\tau}g_{\nu\sigma}+(-2k-q)_{\nu}g_{\sigma\tau}+(2q+k)_{\sigma}g_{\nu\tau}]
\frac{P^{{\rm FD}\,\lambda\sigma}(k)}{k^2} 
\frac{P^{{\rm FD}\,\rho\tau}(k+q)}{(k+q)^2} , \label{eq11}
\\
i\delta^{ab}\Pi^{(c)}_{\mu\nu}(q) =& f^{cad}f^{dbc}g^2 
\int \frac{d^D k}{(2\pi)^D} \frac{n_{\mu}(k)n_{\nu}(k+q)}
{(n(k)\cdot k) (n(k+q)\cdot (k+q))} , \label{eq12}
\end{align}
respectively. 
Here $f^{acd}f^{bcd}=-f^{cad}f^{dbc}=C_A\delta^{ab}$ with $C_A=N_c=3$. 
Since we use the dimensional regularization ($D=4-2\epsilon$), all tadpole 
contributions with massless fields vanish and are not shown.  

Here we comment on the singularity of the FD gauge propagators (\ref{gaugepropagator}). 
As in the covariant gauges, 
the pole from $1/q^2$ at $q^2=0$ should be shifted by the replacement 
$1/q^2\to 1/(q^2+i0)$. 
There is also a singularity from $1/n(q)\cdot q=1/(|q^0|+|\vec{q}\,|)$. 
However, this singularity occurs only at a point $q^{\mu}=0$ in the $D$-dimensional 
phase space and does not need the $+i0$ prescription.

For calculation, 
we split the FD gauge gluon propagators $iP^{\rm FD}_{\mu\nu}(q)/q^2$ in eq.\,(\ref{eq11})
into two parts, $-ig_{\mu\nu}/q^2$ (``$g$'', Feynman gauge propagator) 
and $i(n_{\mu}(q)q_{\nu}+q_{\mu}n_{\nu}(q))/(n(q)\cdot q)q^2$ (``$n$''). 
Eq.~(\ref{eq11})  is then divided as 
\begin{eqnarray}
\Pi^{(b)}=\Pi^{(b, gg)}+\Pi^{(b, gn)}+\Pi^{(b, nn)}.
\end{eqnarray}

The $(gg)$ part, $\Pi^{(b, gg)}_{\mu\nu}(q)$, is 
the self energy in the Feynman gauge. As is well known, its 
UV singular term is~\cite{Srednicki:2007qs}. 
\begin{align}
i\Pi_{\mu\nu}^{(b, gg)}(q)|_{div}  &= -\frac{i}{2}\frac{C_A g^2}{(4\pi)^2\epsilon}
\left( -\frac{19}{6}q^2g_{\mu\nu} + \frac{11}{3}q_{\mu}q_{\nu} \right) 
\nonumber \\
&= -\frac{i}{2}\frac{C_A g^2}{(4\pi)^2\epsilon} 
\left[ \frac{1}{2}(q^0)^2+\frac{19}{6}|\vec{q}\,|^2 , \;  
-\frac{11}{3}q^0{q}^j , \;  
\frac{19}{6}q^2\delta^{ij}+\frac{11}{3}q^i q^j  \right] . 
\label{selfgg}
\end{align}
In the second line, we show 
$i\Pi_{00}$, $i\Pi_{0j}$, and $i\Pi_{ij}$ for later convenience. 

We next evaluate $i\Pi_{\mu\nu}^{(b,gn)}$. It contains 
loop integrals with a factor of $n(k)\cdot k=|k^0|+|\vec{k}\,|$ 
in the denominator, such as 
\begin{equation}
\int \frac{d^D k}{(2\pi)^D} \frac{ 1 }{ 
(|k^0|+|\vec{k}\,|) (k+q)^2 }
\left( {\rm sgn}(k^0), \; -\frac{{k}^j}{|\vec{k}\,|} \right) . 
\label{sampleintegral}
\end{equation}
Here the momentum integration is to be understood 
as $d^D k=d(k^0)d^{D-1}\vec{k}$, 
namely in $(D-1)$-dimensional space and 1-dimensional time. 

Since $|k^0|+|\vec{k}\,|$ is not a polynomial of 
the loop momentum $k^{\mu}$, Feynman's formula to combine 
the denominator of eq.~(\ref{sampleintegral}) into the form 
$((k')^2-C)^n$ does not work. Fortunately, by dimension counting, 
we find that the all UV divergences of 
the integrals like eq.~(\ref{sampleintegral}) are 
polynomials of the components 
of the external momentum $q^{\mu}$. 
We therefore differentiate the integrands in (\ref{eq11}) and (\ref{eq12}) by $q$ to the second 
order, and perform integration of the resulting formulas at 
$q^{\mu}=0$. Details of the integrations are given in Appendix~\ref{sec:AppA}. 

By using the techniques outlined in Appendix~\ref{sec:AppA}, 
the $(gn)$ part of the integral (\ref{eq11}) is found to give 
the following UV divergence, 
\begin{align}
i\Pi_{[00, 0j, ij]}^{(b,gn)}(q)|_{div} & 
= -\frac{i}{2}\frac{C_A g^2}{(4 \pi)^2 \epsilon } 
\left[ 
\frac{20i}{3\pi}(q^0)^2+\left( 2-\frac{76i}{9\pi} \right) |\vec{q}\,|^2 , \;  
-\left( 2-\frac{16i}{9\pi} \right) q^0{q}^j , \right.  
\nonumber \\
& \left. \left( 2+\frac{20i}{9\pi} \right) (q^0)^2\delta^{ij} + 
\left( -2-\frac{148i}{45\pi} \right) |\vec{q}\,|^2\delta^{ij}
+\left( 2+\frac{64i}{45\pi} \right) {q}^i{q}^j  \right] .
\label{selfgn} 
\end{align}
Eq.\,(\ref{selfgn}) has terms with an extra factor of $i/\pi$ 
compared to conventional contributions in the Feynman gauge part (\ref{selfgg}). 
They arise from the UV singular integrals with the $1/(n\cdot k)$ factor, 
which has no on-shell pole. 

The $(nn)$ part of the gluon self energy $i\Pi_{\mu\nu}^{(b,nn)}$ and the 
FP ghost contribution $i\Pi_{\mu\nu}^{(c)}$ are 
evaluated in the same manner. We find 
\begin{align}
i\Pi_{[00,0j,ij]}^{(b,nn)}(q)|_{div} &= -\frac{i}{2}\frac{C_A g^2}{(4 \pi)^2 \epsilon } 
\left[ -\frac{1}{2}(q^0)^2 +\left( \frac{13}{6}-\frac{8i}{\pi} \right) 
|\vec{q}\,|^2 ,  \right. 
\nonumber \\
& \left. 
-\left( \frac{1}{3}-\frac{8i}{3\pi} \right) q^0{q}^j , \;  
\left( -\frac{1}{2}+\frac{8i}{3\pi} \right) q^2\delta^{ij} 
+\left( -1 +\frac{8i}{3\pi} \right) {q}^i{q}^j 
\right] ,
\label{selfnn}
\end{align}
and 
\begin{align}
i\Pi_{[00,0j,ij]}^{(c)}(q)|_{div}= & 
-\frac{i}{2}\frac{C_A g^2}{(4 \pi)^2 \epsilon } 
\left[ -\frac{20i}{3\pi}(q^0)^2 +\frac{4i}{9\pi}|\vec{q}\,|^2 , \;  
\frac{8i}{9\pi}q^0{q}^j ,  \right. 
\nonumber \\
& \left. 
\frac{4i}{9\pi}(q^0)^2\delta^{ij}+\frac{28i}{45\pi}|\vec{q}\,|^2\delta^{ij}
+\frac{56i}{45\pi} {q}^i{q}^j 
\right] , 
\label{selffp}
\end{align}
respectively. 
In contrast to the light-cone gauge~\cite{Leibbrandt:1987qv} where $n^{\mu}$ is 
a constant vector, the FP ghost contribution $i\Pi^{(c)}$ does not vanish. 
Because the ghost loop in the FD gauge has no on-shell pole, 
there is no term without a factor of $i/\pi$ in (\ref{selffp}). 

Summing eqs.~(\ref{selfgg}, \ref{selfgn}, \ref{selfnn}, \ref{selffp}), 
the gluon self energy in the FD gauge is 
\begin{equation}
i\Pi_{[00,0j,ij]}^{{\rm FD}(b+c)}(q)|_{div} =
-\frac{iC_A g^2}{(4 \pi)^2 \epsilon } 
\left[ \left( \frac{11}{3} -\frac{8i}{\pi} \right) |\vec{q}\,|^2 , \; 
-\left( 3-\frac{8i}{3\pi} \right) q^0{q}^j , \;  
\left( \frac{7}{3}+\frac{8i}{3\pi} \right) (q^2\delta^{ij}+{q}^i{q}^j) 
\right] ,
\label{selfenergy}
\end{equation}
or, equivalently, 
\begin{align}
i\Pi_{\mu\nu}^{{\rm FD}(b+c)}(q)|_{div} = &
-\frac{iC_A g^2}{(4 \pi)^2 \epsilon } 
\left[ 
\left( \frac{7}{3} +\frac{8i}{3\pi} \right) (-q^2 g_{\mu\nu}+q_{\mu}q_{\nu})  
\right. 
\nonumber \\ 
& \left. 
+\left( \frac{2}{3} -\frac{16i}{3\pi} \right) 
( q^0(q_{\mu}t_{\nu} +t_{\mu}q_{\nu} ) -2q^2 t_{\mu}t_{\nu} ) \right]  .
\label{selfenergy2} 
\end{align}
We observe that $q^{\mu}\Pi_{\mu\nu}^{\rm FD}(q)|_{div}\neq 0$ but 
$q^{\mu}q^{\nu}\Pi_{\mu\nu}^{\rm FD}(q)|_{div}=0$. 
In fact,  $q^{\mu}q^{\nu}\Pi_{\mu\nu}^{\rm FD}(q)=0$ also holds for 
the UV finite part. 

By substituting the self energy (\ref{selfenergy2}) 
into the (b+c) diagram correction to the amplitude (\ref{ampbc}), 
we find 
\begin{align}
i{\cal M}^{(b+c)}|_{div} & = 
i\frac{C_A g^4}{(4\pi)^2\epsilon}
\bar{v}(p_2)(T^a)_{ji} \gamma^{\mu}u(p_1) \frac{1}{q^2}
\bar{u}(p_3)(T^a)_{lm}\gamma^{\nu} v(p_4) 
\left[ \left( -\frac{7}{3} -\frac{8i}{3\pi}\right) (-g_{\mu\nu}+t_{\mu}t_{\nu}) 
\right. 
\nonumber \\
& \left. + \left( \frac{11}{3}-\frac{8i}{\pi} \right) t_{\mu}t_{\nu} 
+\left(-\frac{4}{3}+\frac{32i}{3\pi} \right) \frac{|q^0|}{|\vec{q}\,|} 
t_{\mu}t_{\nu} \right] .  
\label{dmself}
\end{align}

Next, we calculate 
the UV-divergent parts of the vertex corrections (d, e) to the 
$q_i(p_1)\bar{q}_j(p_2)\to g^a(q)$ vertex, as well as the 
wave function correction (f) of external quarks. 

First, the $(g,g,q)$ loop contribution (d) is 
\begin{align}
& \bar{v}(p_2)(T^a)_{ji}i\Gamma^{(d)\mu}(-q,p_1,p_2)u(p_1)   
\nonumber \\
& =  
i g^3 f^{acd} (T^c T^d)_{ji} 
\int \frac{d^D k}{(2\pi)^D} \frac{1}{ k^2 (k+q)^2(k+p_2)^2}
\nonumber \\
&\quad \times [ (-k+q)^{\rho}g^{\mu\lambda} +(2k+q)^{\mu}g^{\lambda\rho}
-(k+2q)^{\lambda}g^{\mu\rho} ] 
\nonumber \\
&\quad \times P^{{\rm FD}}_{\sigma\lambda}(k)P^{{\rm FD}}_{\rho\tau}(k+q)
\bar{v}(p_2)\gamma^{\sigma}(-\ksla-\psla_2) \gamma^{\tau}
u(p_1) . \label{eq31}
\end{align}
Here $if^{acd}(T^c T^d)_{ji}=-\frac{1}{2}C_A(T^a)_{ji}$. 
By dimension counting, 
the UV divergent part of (\ref{eq31}) should be independent of 
the external momenta $(q,p_1,p_2)$. 

Again, we split the gluon propagators in eq.~(\ref{eq31}) into 
``$g$'' and ``$n$'' parts.  
The Feynman gauge $(gg)$ part is 
\begin{equation}
i\Gamma^{(d, gg)\mu}|_{div}= i\frac{C_A g^3}{(4\pi)^2\epsilon}
\left( -\frac{3}{2} \gamma^{\mu} \right), 
\label{vertexgg}
\end{equation}
The other parts, $(gn)$ and $(nn)$, are 
\begin{equation}
i\Gamma^{(d, gn)[0,i]}|_{div} = i\frac{C_A g^3}{(4\pi)^2\epsilon}
\left[ \left( \frac{3}{2}-\frac{2i}{\pi}\right)\gamma^0, \; 
\left( \frac{3}{2} +\frac{2i}{3\pi} \right) \gamma^i \right] ,
\label{vertexgn}
\end{equation}
and 
\begin{equation}
i\Gamma^{(d,nn)[0,i]}|_{div} = i\frac{C_A g^3}{(4\pi)^2\epsilon}
\left[ \left( \frac{1}{2}-\frac{2i}{\pi}\right)\gamma^0, \; 
\left( -\frac{1}{6} +\frac{2i}{3\pi} \right) \gamma^i \right] ,
\label{vertexnn}
\end{equation}
respectively. Their summation then gives 
\begin{equation}
i\Gamma^{(d)[0,i]}|_{div} = i\frac{C_A g^3}{(4\pi)^2\epsilon}
\left[ \left( \frac{1}{2}-\frac{4i}{\pi}\right)\gamma^0, \; 
\left( -\frac{1}{6} +\frac{4i}{3\pi} \right) \gamma^i \right] .
\label{vertexd}
\end{equation}

The $(q,q,g)$ loop contribution (e) is given by 
\begin{align}
& \bar{v}(p_2)(T^a)_{ji} i\Gamma^{(e)\mu}(-q,p_1,p_2) u(p_1)  
\nonumber \\
& = 
g^3 (T^c T^a T^c)_{ji} 
\int \frac{d^D k}{(2\pi)^D} \frac{1}{k^2(k+p_1)^2(k-p_2)^2}
\nonumber \\
&\quad \times P^{\rm FD}_{\nu\rho}(k) 
\bar{v}(p_2) 
\gamma^{\nu}(\ksla-\psla_2) \gamma^{\mu} (\ksla+\psla_1) \gamma^{\rho} u(p_1) . \label{eq35}
\end{align}
Here $(T^c T^a T^c)_{ji}=(C_F-\frac{1}{2}C_A)(T^a)_{ji}$ with 
$C_F=(N_c^2-1)/(2N_c)=4/3$. 
After splitting the gluon propagator into ``$g$'' and ``$n$'' parts, 
we have 
\begin{subequations}
\begin{equation}
i\Gamma^{(e,g)\mu} |_{div}  =i\frac{g^3}{(4\pi)^2\epsilon} (C_A-2C_F)
\left[ \frac{1}{2}\gamma^{\mu}  \right] , \label{vertexeg}
\end{equation}    
\begin{equation}
i\Gamma^{(e,n)\mu} |_{div}  =i\frac{g^3}{(4\pi)^2\epsilon} (C_A-2C_F)
\left[ -\gamma^{\mu} \right] .  
\label{vertexen}
\end{equation}
\end{subequations}
Note that the ``$n$'' part (\ref{vertexen}) is Lorentz covariant, unlike 
the cases of the corrections $(b,c,d)$. 

We further include the contribution from the quark wave function correction (f) 
to the vertex $i\Gamma^{\rm FD}$. 
The quark self energy in the FD gauge is 
\begin{equation}
i\Sigma_q^{\rm FD}(p_i) = g^2 C_F \int \frac{d^D k}{(2\pi)^D}
\frac{\gamma^{\nu}(\ksla+\psla_i)\gamma^{\rho}}{k^2(k+p_i)^2} P_{\nu\rho}^{\rm FD}(k) .
\end{equation}
Its UV divergence is, after splitting $P_{\mu\nu}^{\rm FD}$ into $O(g_{\mu\nu})$ and $O(nk)$ terms, 
\begin{equation}
(i\Sigma_q^{(g)}(p_i)|_{div} , \; i\Sigma_q^{(n)}(p_i)|_{div}  )
= i\frac{g^2 C_F}{(4\pi)^2\epsilon}  \left(  \psla_i, \;  - 2\psla_i  \right)  .
\end{equation}
Then   
\begin{align}
i\Gamma^{(f,g)\mu} |_{div} & =i\frac{g^3}{(4\pi)^2\epsilon} C_F 
\left[ \gamma^{\mu}  \right] , \label{vertexfg}
\\
i\Gamma^{(f,n)\mu} |_{div} & =i\frac{g^3}{(4\pi)^2\epsilon} C_F 
\left[ -2\gamma^{\mu} \right] .  \label{vertexfn}
\end{align}
They exactly cancel the $O(C_F)$ contributions of the $(q,q,g)$ vertex correction $i\Gamma^{(e)\mu}|_{div}$ 
(\ref{vertexeg}, \ref{vertexen}). 
In total, the UV-divergent ${q}qg$ vertex correction in the FD gauge is 
\begin{equation}
i\Gamma^{(d+e+f)(0,i)}|_{div} = i\frac{g^3 C_A }{(4\pi)^2\epsilon} 
\left[ -\frac{4i}{\pi}\gamma^0, \; \left( -\frac{2}{3}+\frac{4i}{3\pi} \right) \gamma^i 
\right] . 
\label{vertexdef}
\end{equation}

The correction to the amplitude by $i\Gamma^{(d+e+f)}$ 
for the initial ${q}qg$ vertex is, by using Eq.~(\ref{mvertex3}), 
\begin{align}
i{\cal M}^{(d+e+f)}_{init}|_{div} & =  i\frac{C_A g^4}{(4\pi)^2\epsilon} 
\bar{v}(p_2)(T^a)_{ji}\gamma^{\mu}u(p_1) \frac{1}{q^2}\bar{u}(p_3)(T^a)_{lm}
\gamma^{\nu} v(p_4)
\nonumber \\
& \times \left[ \left( -\frac{2}{3}+\frac{4i}{3\pi} \right) 
(-g_{\mu\nu}+t_{\mu}t_{\nu}) + \left( \frac{2}{3}-\frac{16i}{3\pi} \right) 
\frac{|q^0|}{|\vec{q}\,|} t_{\mu}t_{\nu} +\frac{4i}{\pi} t_{\mu} t_{\nu} \right] .
\label{dmvertex} 
\end{align}
The 
final ${q}'q'g$ vertex correction $i{\cal M}^{(d+e+f)}_{fin}|_{div}$ is identical to eq.~(\ref{dmvertex}). 

Finally, the box corrections $\Delta{\cal M}^{(g,h)}$ are, 
as in the covariant gauges, UV finite. 

In total, UV-divergent part of the gluon loop corrections to the amplitude is 
\begin{align} 
i{\cal M}^{\rm FD}({\rm corr})|_{div} & =
i\frac{C_A g^4}{(4\pi)^2\epsilon} 
\bar{v}(p_2)(T^a)_{ji} \gamma^{\mu}u(p_1) \frac{1}{q^2}
\bar{u}(p_3)(T^a)_{lm} \gamma^{\nu} v(p_4)
\nonumber \\
& \times 
\left[  -\frac{11}{3}  
(-g_{\mu\nu}+t_{\mu}t_{\nu})  +\frac{11}{3} t_{\mu} t_{\nu} \right] 
\nonumber \\
& = i{\cal M}^{(a)} \times \left( 
\frac{11}{3}\frac{C_A g^2}{(4\pi)^2\epsilon} \right) .
\end{align}
This result is identical to the one in the covariant gauges 
and consistent with the beta function~\cite{Gross:1973id,Politzer:1973fx}
$\beta(g)=-\frac{11}{3}C_A g^3/(4\pi)^2$ of the gauge coupling $g$. 
This result gives an evidence that the FD gauge fixing (\ref{FDgaugefixing0}) 
with the gauge vector (\ref{FDlightcone}) in the momentum space 
gives
a consistent procedure for gauge fixing. 


\subsection{Transverse and longitudinal contributions}

We have seen that loop corrections in the FD gauge have unconventional 
UV divergences whose coefficients differ from the conventional 
ones by a factor of $O(i/\pi)$. 
For better understanding of this type 
of the loop contributions, 
we examine the contributions of 
the transverse and longitudinal parts of the off-shell gluons 
separately in this subsection. 

The FD gauge polarization tensor $P^{{\rm FD}\mu\nu}(k)$ 
is decomposed into the transverse part 
$P_T^{\mu\nu}$ and the longitudinal part $P_L^{\mu\nu}$, 
as \cite{Hagiwara:2020tbx} 
\begin{align}
P^{{\rm FD}\mu\nu}(k) 
& = P_T^{\mu\nu}(k) + P_L^{\mu\nu}(k) 
\nonumber \\
&= \delta^{\mu}_i\delta^{\nu}_j \left( 
\delta^{ij} -\frac{{k}^{i}{k}^{j}}{|\vec{k}\,|^2} 
\right) +  k^2 \frac{n^{\mu}(k) n^{\nu}(k)} {(n(k)\cdot k)^2}  . \label{eq9}
\end{align}
This equation can be verified by using the explicit form of 
$n^{\mu}(k)$ (\ref{FDlightcone}). 
The gluon propagator is then decomposed as 
\begin{equation}
iG^{{\rm FD}\mu\nu}(k) = i\frac{P_T^{\mu\nu}(k)}{k^2+i0} + 
i\frac{n^{\mu}(k) n^{\nu}(k)} {(n(k)\cdot k)^2} .  \label{eq10}
\end{equation}
Since $1/(n(k)\cdot k)=1/(|k^0|+|\vec{k}\,|)$ diverges 
only at a point $k^{\mu}=0$, the longitudinal part 
of the propagator does not correspond to physical states. 

In this subsection, we separate the UV-divergent one-loop gluon 
corrections to 
the $q\bar{q}\to g(p)\to q'\bar{q}'$ amplitude into transverse (T) and 
longitudinal (L) internal gluons. For simplicity, 
we work in the center-of-mass frame of $q\bar{q}$, 
where $q^{\mu}=(Q,\vec{0})$ ($Q>0$).\footnote{The case of 
general $q^{\mu}$ is briefly discussed in Appendix B.}
Note that, in this case, we have 
$n^{\mu}(q)=(1,\vec{n})$ where $\vec{n}=-\vec{q}/|\vec{q}\,|$ is a 
unit 3D vector whose direction is not determined in the $|\vec{q}\,|\to 0$ limit. 
We will find, nevertheless, that this ambiguity of $n^{\mu}(q)$ 
does not affect the amplitude (\ref{eq4}). 

We start from the gluon self energy. The transverse-transverse (TT), 
transverse-longitudinal (TL), and longitudinal-longitudinal (LL) parts are, respectively,  
\begin{subequations}
\begin{align}
i\Pi^{(bTT)}_{[00,ij]}(q)|_{div}  
& = C_A g^2 \frac{i}{(4\pi)^2 \epsilon} Q^2 \left[ 0, \;  
\frac{1}{3}\delta^{ij}  \right] , \label{eq16}
\\
%
i\Pi^{(bTL)}_{[00,ij]}(q)|_{div}  
& = C_A g^2 \frac{i}{(4\pi)^2\epsilon} Q^2 
\left[ 0, \;  -\frac{8}{3}\delta^{ij}  \right] ,  \label{eq21}
\\
%
i\Pi^{(bLL)}_{[00,ij]}(q)|_{div}  
& = C_A g^2 \frac{i}{(4\pi)^2\epsilon}Q^2
\left[ -\frac{10i}{3\pi} , 
 -\frac{22i}{9\pi}  \delta^{ij} 
 \right], \label{eq25}
\end{align}
\end{subequations}
while $i\Pi_{0j}(q)=0$ by space rotational invariance. 
It is seen that 
the unconventional $O(i/\pi)$ term in $i\Pi^{\rm FD}$ arises from the (LL) part 
(\ref{eq25}), 
where the intermediate propagators (two longitudinal gluons) have no cuts. 
%
The FP ghost contribution for $q^{\mu}=(Q,\vec{0})$ is, 
from eq.~(\ref{selffp}), 
\begin{equation}
i\Pi^{(c)}_{[00, ij]}(q)|_{div}  
=  C_A g^2 \frac{i}{(4\pi)^2\epsilon}Q^2 
\left[   \frac{10i}{3\pi}  ,  \;  
-\frac{2i}{9\pi} \delta^{ij} 
 \right]  . \label{eq28}
\end{equation}
In total, the gluon self energy is 
\begin{equation}
i\Pi^{(b+c)}_{[00,ij]}(q)|_{div} =  
C_A g^2 \frac{i}{(4\pi)^2\epsilon} Q^2
\left[ 0, \; 
\left( -\frac{7}{3} - \frac{8i}{3\pi}  \right) \delta^{ij} \right] . 
\label{eq30} 
\end{equation}
This result is consistent with the result (\ref{selfenergy}) 
for general $q^{\mu}$, as it must be. 
Since $q^{\mu}=(Q,\vec{0})$ here, $q^{\lambda}\Pi^{(b+c)}_{\mu\lambda}(q)=0$ holds and
the $n(q)$-dependent 
contributions in the correction (\ref{eq7}) to the scattering amplitudes vanish. 

The vertex correction (d) is, as for the gluon self energy (b), decomposed 
into (TT), (TL), and (LL) parts as 
\begin{subequations}
\begin{align}
i\Gamma^{(dTT)\mu} |_{div} & =    
i\frac{C_A g^3}{(4\pi)^2\epsilon} 
\left( -\frac{1}{2} \gamma^{\mu} \right) , 
\label{vertexdtt} 
\\
%
i\Gamma^{(dTL)[0,i]}|_{div} & =
i\frac{C_A g^3}{(4\pi)^2\epsilon}  \left[ 0, \; \frac{2}{3}\gamma^i  \right] , 
\\
%
i\Gamma^{(dLL)[0,i]}|_{div} & =
i\frac{C_A g^3}{(4\pi)^2\epsilon}  
\left[ 
\left( 1 -\frac{4i}{\pi} \right) \gamma^0, \;\; 
\left( -\frac{1}{3}+\frac{4i}{3\pi} \right) \gamma^i  \right]  .
\end{align}
\end{subequations}
The unconventional $O(i/\pi)$ term appears only in the (LL) part 
with two unphysical propagators, as in the case of the gluon 
self energy correction (b) given in eqs.~(\ref{eq16}, \ref{eq21}, \ref{eq25}). 

The vertex correction (e) is decomposed by separating the gluon propagator 
in the loop, as 
\begin{subequations}
\begin{equation}
i\Gamma^{(eT)[0,i]} |_{div} =  
i(C_A-2C_F) \frac{g^3}{(4\pi)^2\epsilon} 
\left[ \frac{1}{2} \gamma^0,\;\; -\frac{1}{6} \gamma^i  \right] ,
\label{vertexet}
\end{equation}    
\begin{equation}
i\Gamma^{(eL)[0,i]}|_{div} =  
i(C_A-2C_F) \frac{g^3}{(4\pi)^2\epsilon}  
\left[ -\gamma^0,\;\; -\frac{1}{3} \gamma^i  \right] . 
\label{vertexel}
\end{equation}
\end{subequations}
giving
\begin{equation}
    i\Gamma^{(e)\mu}|_{div}
    =i(C_A-2C_F)
    \frac{g^3}{(4\pi)^2\epsilon}
    \left(
    -\frac{1}{2}\gamma^\mu\right)
    .
    \label{eq:vexeTeL}
\end{equation}

There are no $O(i/\pi)$ terms in eqs.~(\ref{vertexet}, \ref{vertexel}). 
Likewise, the quark self energy is decomposed as 
\begin{subequations}
\begin{align}
i\Sigma_q^{(T)}(p)|_{div}  &=i \frac{g^2 C_F}{(4\pi)^2\epsilon} \left[  
p^0\gamma^0+\frac{1}{3}{p}^{\,i}\gamma^i  \right] ,
\label{selfqt}
\\
i\Sigma_q^{(L)}(p)|_{div}  &= i\frac{g^2 C_F}{(4\pi)^2\epsilon} \left[  
-2p^0\gamma^0 + \frac{2}{3}{p}^{\,i}\gamma^i  \right] .
\label{selfql}
\end{align}
\end{subequations}
Their sum 
\begin{eqnarray}
    i\Sigma_q(p)|_{div} 
=
i\frac{g^2 C_F}{(4\pi)^2\epsilon} \left[-\slashed p\right]
\end{eqnarray}
contributes to the vertex correction term (f) as 
\begin{align}
i\Gamma^{(f)\mu}|_{div} = & \frac{iC_F g^3}{(4\pi)^2\epsilon}
\left[ - \gamma^{\mu} \right] . 
\label{vertexftl}
\end{align}
Eq.~(\ref{vertexftl}) cancels the $O(C_F)$ terms of 
$i\Gamma^{(e)}$ \eqref{eq:vexeTeL}. 
It is worth nothing that the sum of T and L components of the FD gauge propagator gives 
sensible correction to the vertex corrections (e) and the quark self energy correction in (f).
The total vertex correction (d+e+f) is 
\begin{equation}
i\Gamma^{(d+e+f)[0,i]}|_{div} = i\frac{g^3 C_A }{(4\pi)^2\epsilon} 
\left[ -\frac{4i}{\pi}\gamma^0, \; \left( -\frac{2}{3}+\frac{4i}{3\pi} \right) \gamma^i 
\right] , 
\end{equation}
that agrees with the result (\ref{vertexdef}). 

In calculating corrections to the scattering amplitude, 
we cannot use eq.~(\ref{nslaid}) since $|\vec{q}\,|=0$. Instead, by using 
$\bar{v}(p_2)\gamma^0 u(p_1)=\bar{u}(p_3)\gamma^0 v(p_4)=0$ 
from the quark EOM, 
$q^{\mu}\Pi_{\mu\nu}(q)=0$, and $q_{\mu}\Gamma^{\mu}\propto\gamma^0$, we find 
\begin{equation}
i{\cal M}^{(b+c)}|_{div}  = 
i\frac{C_A g^4}{(4\pi)^2\epsilon}
\bar{v}(p_2)(T^a)_{ji} \gamma^{i}u(p_1) \frac{1}{Q^2}
\bar{u}(p_3)(T^a)_{lm}\gamma^{i} v(p_4) \times 
\left( -\frac{7}{3} -\frac{8i}{3\pi}\right) ,
\label{dmselfcm}
\end{equation}
from the gluon self energy, and 
\begin{equation}
i{\cal M}^{(d+e+f)}|_{div}  =  i\frac{C_A g^4}{(4\pi)^2\epsilon} 
\bar{v}(p_2)(T^a)_{ji}\gamma^j u(p_1) \frac{1}{Q^2}\bar{u}(p_3)(T^a)_{lm}
\gamma^j v(p_4)
\times  \left( -\frac{4}{3}+\frac{8i}{3\pi} \right) ,
\label{dmvertexcm} 
\end{equation}
from the initial and final vertex corrections, respectively. 
Both (\ref{dmselfcm}) and (\ref{dmvertexcm}) are 
independent of $n^{\mu}(q)$, especially of its undetermined space 
components $n^i(q)$. 
The UV-divergent part of the gluon loop corrections to the amplitude is, 
in total, 
\begin{align} 
i{\cal M}^{\rm FD}({\rm corr})|_{div} & =
i\frac{C_A g^4}{(4\pi)^2\epsilon} 
\bar{v}(p_2)(T^a)_{ji} \gamma^j u(p_1) \frac{1}{Q^2}
\bar{u}(p_3)(T^a)_{lm} \gamma^j v(p_4)
\times \left(  -\frac{11}{3} \right)  
\nonumber \\
& = i{\cal M}^{(a)} \times \left( 
\frac{11}{3}\frac{C_A g^2}{(4\pi)^2\epsilon} \right) .
\end{align}
This result is again identical to the one 
in the covariant gauges 
for $q^{\mu}=(Q,\vec{0})$. 


\subsection{Equivalence of the amplitudes in the FD and Feynman gauges}

Up to now, we have only considered the UV-divergent parts of the gluon loop 
corrections. However, on-shell amplitudes in gauge theories should be 
independent of the gauge fixing methods. In this subsection 
we show how all the $n$-dependent terms of the loop correction 
(\ref{eq4}) to the $q\bar{q}\to g\to q'\bar{q}'$ process 
in the FD gauge cancel among each other, including finite parts, 
to leave the amplitudes the same as in the Feynman gauge. 
Here we work on the level of the integrands, without
explicit evaluation of loop integration. 


We start from the box diagrams (g, h) in Fig.~\ref{fig:diagram_qqqq}. 
The contribution from (g) is 
\begin{align}
i{\cal M}^{(g)} & = 
g^4(T^aT^b)_{ji} (T^bT^a)_{lm} \int \frac{d^D k}{(2\pi)^D}
\frac{1}{k^2(k+q)^2(k+p_2)^2(k+p_4)^2}
\nonumber \\
&\quad \times \bar{v}(p_2)\gamma^{\mu}(-\ksla-\psla_2)\gamma^{\lambda}u(p_1)
\cdot \bar{u}(p_3)\gamma^{\sigma}(-\ksla-\psla_4)\gamma^{\nu}v(p_4) 
\nonumber \\
&\quad \times P^{\rm FD}_{\mu\nu}(k) P^{\rm FD}_{\lambda\sigma}(k+q) . 
\label{dmboxg0}
\end{align}
Eq.~(\ref{dmboxg0}) is UV finite and has not been discussed in the previous subsections. 

Now we focus on the $n$-dependent parts of eq.~(\ref{dmboxg0}), which give 
the difference between the Feynman and the FD gauges. 
It is seen that the gluon momenta in the $n$-dependent parts of 
the gluon propagators cancel the attached quark propagators, 
or `pinch', 
reducing the kinematic structure to that of the vertex or 
gluon self energy contributions~\cite{Cornwall:1989gv,Binosi:2009qm}. 
For example, $k_{\mu}n_{\nu}(k)$ part of 
$P^{\rm FD}_{\mu\nu}(k)$ in eq.~(\ref{dmboxg0}) reduces the integrand as, 
by using the EOMs for external quarks, 
\begin{align}
& \bar{v}(p_2)\gamma^{\mu} \frac{ -\ksla-\psla_2 }{(k+p_2)^2}\gamma^{\lambda}u(p_1)
\cdot (k_{\mu}n_{\nu}(k))
\cdot \bar{u}(p_3)\gamma^{\sigma}
\frac{ -\ksla-\psla_4 }{(k+p_4)^2} \gamma^{\nu}v(p_4)
\nonumber \\
& = 
 \bar{v}(p_2) \ksla \frac{ -\ksla-\psla_2 }{(k+p_2)^2}
\gamma^{\lambda}u(p_1)\cdot \bar{u}(p_3)\gamma^{\sigma}
\frac{ -\ksla-\psla_4 }{(k+p_4)^2} \nsla(k) v(p_4)
\nonumber \\
& =  
- \bar{v}(p_2) \gamma^{\lambda}u(p_1)\cdot \bar{u}(p_3)\gamma^{\sigma}
\frac{ -\ksla-\psla_4 }{(k+p_4)^2} \nsla(k) v(p_4) ,
\label{pinch1}
\end{align}
times $P^{\rm FD}_{\lambda\sigma}(k+q)/[k^2(n(k)\cdot k)(k+q)^2]$. 
The last line of eq.~(\ref{pinch1}) is independent of $p_2$, 
giving a contribution with the kinematic structure of the vertex correction to the 
final ${q}'q'g$ coupling. 

After successively applying the `pinch' method, 
the $(gn)$ and $(nn)$ parts of the box contributions 
$i{\cal M}^{(g+h)}$ can be expressed as 
\begin{equation}
 i{\cal M}^{(g+h,gn+nn)} =  i{\cal M}^{box}_1+i{\cal M}^{box}_2+i{\cal M}^{box}_3 , 
\end{equation}
where 
\begin{subequations}
\begin{align}
 i{\cal M}^{box}_1= & -\frac{1}{2}
C_A g^4 (T^a)_{ji} (T^a)_{lm} 
\bar{u}(p_3)\gamma_{\mu}v(p_4) 
\int \frac{d^Dk}{(2\pi)^D} \frac{1}{k^2 (k+q)^2 (k+p_2)^2} 
\nonumber 
\\
& \times   \bar{v}(p_2)
\left\{ 
\frac{ 
\gamma^{\mu}(\ksla+\psla_2)\nsla(k+q)   } 
{n(k+q)\cdot(k+q)}  
 + \frac{ \nsla(k) (\ksla+\psla_2)\gamma^{\mu}  }
{n(k)\cdot k }  
\right. 
\nonumber 
\\
& \left. 
-\frac{  k^{\mu} \nsla(k) (\ksla+\psla_2) \nsla(k+q)  } 
{  (n(k)\cdot k)(n(k+q)\cdot(k+q)) } 
\right\}  u(p_1)  ,
 \label{eq:iMbox1}
 \end{align}    
\begin{align}
 i{\cal M}^{box}_2= & -\frac{1}{2}
C_A g^4 (T^a)_{ji} (T^a)_{lm} \bar{v}(p_2)\gamma_{\mu}u(p_1) 
\int \frac{d^Dk}{(2\pi)^D} \frac{1}{k^2 (k+q)^2 (k+p_3)^2} 
\nonumber \\
& \times  \bar{u}(p_3) \left\{ 
\frac{ \gamma^{\mu}(\ksla+\psla_3)\nsla(k+q) } {n(k+q)\cdot(k+q)} 
+ \frac{ \nsla(k) (\ksla+\psla_3)\gamma^{\mu}  } {n(k)\cdot k} 
\right.
\nonumber \\
& \left. 
-
\frac{  k^{\mu} \nsla(k) (\ksla+\psla_3) \nsla(k+q)  }  
{ (n(k)\cdot k)(n(k+q)\cdot(k+q)) } 
\right\} v(p_4)   ,
\label{eq:iMbox2}
\end{align}    
\begin{align}
 i{\cal M}^{box}_3= & \frac{1}{2}
C_A g^4 (T^a)_{ji} (T^a)_{lm} 
\bar{v}(p_2)\gamma^{\mu}u(p_1) \cdot 
\bar{u}(p_3)\gamma^{\nu}v(p_4) 
\nonumber \\
& \times  \int \frac{d^Dk}{(2\pi)^D} \frac{ 1}{k^2 (k+q)^2} 
\frac{ 
n_{\mu}(k+q) n_{\nu}(k) + n_{\mu}(k) n_{\nu}(k+q) }
{(n(k)\cdot k)(n(k+q)\cdot(k+q))}    .
\label{pinch2}
\end{align}
\end{subequations}
Examining the dependence of $i{\cal M} ^{box}_{1-3}$ in \eqref{eq:iMbox1},\eqref{eq:iMbox2},\eqref{pinch2} on 
the external momenta $q$ and $p_i$ ($i=1$ to $4$), we find that these three parts 
kinematically behave as the corrections on the initial $qqg$ vertex, 
on the final $q'q'g$, and on the gluon self energy, respectively. 

Next, we examine the $n$-dependent parts of the vertex correction contributions 
${\cal M}^{(d+e+f)}$, coming from the gluon propagators in the initial $qqg$ and the final $q'q'g$ 
vertex functions $\Gamma^{\mu}$, 
and also the $n(q)$ dependence coming from the FD gauge propagator $iP^{\rm FD}_{\mu\nu}(q)$ in (\ref{eq6}). 

The $(gn,nn)$ parts of the vertex function $i\Gamma^{(d)\mu}$ for 
the initial $qqg$ vertex are written as, after applying the EOMs for 
external quarks, 
\begin{equation}
i\Gamma^{(d, gn+nn)\mu} = 
i\Gamma^{(d)\mu}_1 + i\Gamma^{(d)\mu}_2 +i\Gamma^{(d)\mu}_3 , 
\label{iGamma123}
\end{equation}
where
\begin{subequations}
\begin{align}
i\Gamma^{(d)\mu}_1 = & -\frac{1}{2}C_A g^3 \int \frac{d^D k}{(2\pi)^D} 
\frac{1}{k^2(k+q)^2(k+p_2)^2} 
\nonumber \\
& \times \left[ 
\frac{\nsla(k)(\ksla+\psla_2) 
(q^{\mu} \ksla + q^2 \gamma^{\mu} ) }
{n(k)\cdot k}  \right.  
\nonumber \\ 
& 
+ \frac{ (-q^{\mu} (\ksla+\qsla) + q^2 \gamma^{\mu} )
(\ksla+\psla_2)\nsla(k+q) } {n(k+q)\cdot(k+q)} 
\nonumber \\ 
& \left. 
+\frac{ -k^{\mu} q^2 +q^{\mu}(q\cdot k)} {(n(k)\cdot k)(n(k+q)\cdot (k+q))}
\nsla(k)(\ksla+\psla_2)\nsla(k+q) \right] , 
\label{eq:iGamma1}
\end{align}    
\begin{align}
i\Gamma^{(d)\mu}_2 = &  -\frac{1}{2}C_A g^3 \int \frac{d^D k}{(2\pi)^D} 
\left[ -\frac{\nsla(k)(\ksla+\psla_2)\gamma^{\mu} } {k^2(n(k)\cdot k)(k+p_2)^2} 
-\frac{ \gamma^{\mu} (\ksla+\psla_2)\nsla(k+q) }
 {(k+q)^2(n(k+q)\cdot (k+q))(k+p_2)^2}   \right] , 
\label{eq:iGamma2}
\end{align}    
\begin{align}
i\Gamma^{(d)\mu}_3 = & -\frac{1}{2}C_A g^3 \int \frac{d^D k}{(2\pi)^D} 
\frac{1}{k^2(k+q)^2} 
\nonumber \\
& \times \left[  \frac{1}{ n(k)\cdot k }\left( 
-n(k)\cdot(2q+k)\gamma^{\mu} -n^{\mu}(k)\ksla +(3k+q)^{\mu}\nsla(k) \right) 
\right. 
\nonumber \\
& 
+ \frac{1}{ n(k+q)\cdot (k+q) } \left( 
-n(k+q)\cdot(k-q)\gamma^{\mu} -n^{\mu}(k+q)\ksla +(3k+2q)^{\mu}\nsla(k+q) \right) 
\nonumber \\
& 
+\frac{1}{(n(k)\cdot k)(n(k+q)\cdot(k+q))}  \left\{ 
\left( (k^2-q^2)n^{\mu}(k)-k^{\mu}(n(k)\cdot k)+q^{\mu}(n(k)\cdot q) \right) 
\nsla(k+q)  \right. 
\nonumber \\
& + \left( (k^2+2q\cdot k)n^{\mu}(k+q)-k^{\mu}(n(k+q)\cdot (k+q))
-q^{\mu}(n(k+q)\cdot k) \right) 
\nsla(k)
\nonumber \\ 
& \left. \left. 
+\left ( n^{\mu}(k)(k-q)\cdot n(k+q) +n^{\mu}(k+q)(2q+k)\cdot n(k) 
-(2k+q)^{\mu} n(k)\cdot n(k+q) \right)  \ksla \right\} \right] .
\label{pinch3}
\end{align}
\end{subequations}
Note that the integral (\ref{eq:iGamma1}) depends on 
both $q$ and $p_2$, the first term in (\ref{eq:iGamma2}) 
depends only on $p_2$, whereas the second term depends 
only on $p_1$, after transforming $k\to k+q$. 
The integrals in (\ref{pinch3}) depend only on $q$. 

The $O(n)$ contribution from 
the vertex function $i\Gamma^{(e)\mu}$ for 
the initial $q{q}g$ vertex is 
\begin{align}
i\Gamma^{(e,n)\mu} = & (C_F - \frac{1}{2} C_A) g^3 \int \frac{d^D k}{(2\pi)^D} 
\frac{1}{k^2(n(k)\cdot k)} \left[ 
\frac{ \gamma^{\mu}(\ksla+\psla_1)\nsla(k)}{(k+p_1)^2} + 
\frac{ \nsla(k)(\ksla+\psla_2)\gamma^{\mu} }{(k+p_2)^2} \right].
\label{pinch6}
\end{align}
From the $O(n)$ part of the quark self energy 
\begin{equation}
i\Sigma_q^{(n)}(p_i) = C_F g^2 \int \frac{d^D k}{(2\pi)^D} 
\frac{-\psla_i(\ksla+\psla_i)\nsla(k)  -\nsla(k)(\ksla+\psla_i)\psla_i }
{k^2(n(k)\cdot k)(k+p_i)^2} , 
\end{equation}
we obtain 
\begin{equation}
i\Gamma^{(f,n)\mu} = - C_F g^3 \int \frac{d^D k}{(2\pi)^D} 
\frac{1}{k^2(n(k)\cdot k)} \left[ 
\frac{\gamma^{\mu}(\ksla+\psla_1)\nsla(k)}{(k+p_1)^2} 
+\frac{\nsla(k)(\ksla+\psla_2)\gamma^{\mu}}{(k+p_2)^2} 
\right] ,
\label{pinch7}
\end{equation}
for the initial $qqg$ vertex. 
The $O(C_F)$ part of (\ref{pinch6}) is exactly cancelled by 
the quark wave function correction (\ref{pinch7}). 
The remaining $O(C_A)$ part of (\ref{pinch6}) cancels 
$i\Gamma^{(d)\mu}_2$ in eq.~\eqref{eq:iGamma2}, after momentum transformation 
$k\to -k-q$ in some terms. 

In the remaining parts of eq.~(\ref{iGamma123}), only $i\Gamma^{(d)\mu}_1$\,\eqref{eq:iGamma1} has 
$p_2$ dependence. Its contribution to the amplitude is, from eq.~(\ref{eq6}), 
\begin{align}
i{\cal M}^{vert}_{init,1} =  &  \bar{v}(p_2)(T^a)_{ji}i\Gamma^{(d)\mu}_1u(p_1)
\frac{i}{q^2} \left( -g_{\mu\nu} + \frac{q_{\mu} n_{\nu}(q)}{n(q)\cdot q} \right)
(-ig)\bar{u}(p_3)(T^a)_{lm}\gamma^{\nu}v(p_4) 
\nonumber \\ 
= &  i{\cal M}^{vert}_{init,11}+ i{\cal M}^{vert}_{init,12}  ,
\end{align}
where
\begin{subequations}
\begin{align}
i{\cal M}^{vert}_{init, 11} = & 
 -\frac{1}{2}C_A g^4 (T^a)_{ji}(T^a)_{lm} 
 \bar{u}(p_3)\gamma_{\mu}v(p_4) 
 \int \frac{d^D k}{(2\pi)^D} 
\frac{1}{k^2(k+q)^2(k+p_2)^2}
\nonumber \\
& \times \bar{v}(p_2) \left[ 
- \frac{ \nsla(k)(\ksla+\psla_2)\gamma^{\mu}  }
{n(k)\cdot k}  
- \frac{ \gamma^{\mu} (\ksla+\psla_2)\nsla(k+q)  } {n(k+q)\cdot(k+q)}  \right. 
\nonumber \\ 
& \left. 
+\frac{  k^{\mu} \nsla(k)(\ksla+\psla_2)\nsla(k+q) 
 } {(n(k)\cdot k)(n(k+q)\cdot (k+q))}
\right]  u(p_1) ,
\label{eq:iMinit11}
\end{align}    
\begin{align}
i{\cal M}^{vert}_{init,12} =
&  -\frac{1}{2}C_A g^4 (T^a)_{ji}(T^a)_{lm} 
\bar{u}(p_3) \nsla(q) v(p_4) 
\int \frac{d^D k}{(2\pi)^D} \frac{1}{k^2(k+q)^2}
\nonumber \\
& \times \left[ 
\frac{ \bar{v}(p_2) \nsla(k) u(p_1) }
{n(k)\cdot k}  
+ \frac{ -\bar{v}(p_2) \nsla(k+q) u(p_1)  } {n(k+q)\cdot(k+q)}  \right] 
 . 
\label{pinch8}
\end{align}
\end{subequations}
$i{\cal M}^{vert}_{init,11}$ in \eqref{eq:iMinit11} cancels $i{\cal M}^{box}_1$ in \eqref{eq:iMbox1}, 
while $i{\cal M}^{vert}_{init,12}$ in eq.~\eqref{pinch8} has no $p_2$ dependence in the loops 
and behaves as the gluon self energy correction. 

In the same manner, the $n$-dependent part of the correction 
to the final $q'{q}'g$ vertex cancels  $i{\cal M}^{box}_2$ in \eqref{eq:iMbox2}, 
leaving only the gluon self-energy-like correction, 
which we denote as $i{\cal M}^{vert}_{fin,12}$
Also, $i\Gamma^{(d)\mu}_3$\,\eqref{pinch3} on the initial and final $q{q}g$ vertices 
give self-energy-like contributions to the amplitude, 
which we denote as $i{\cal M}^{vert}_3$. 

Note that the Feynman gauge part of the vertex function, 
$i\Gamma^{(d+e+f, g)\mu}$, does not give $n(q)$-dependent contribution 
because of the relation 
$q_{\mu}\Gamma^{(d+e+f, g)\mu}(q)=0$ for the on-shell external quarks. 

Therefore, all the remaining $n$-dependent box/vertex correction 
parts of the amplitude, $i{\cal M}^{box}_3$ (\ref{pinch2}), 
$i{\cal M}^{vert}_{init,12}$ (\ref{pinch8}), $i{\cal M}^{vert}_{fin,12}$, 
and $i{\cal M}^{vert}_3$ show momentum dependence of 
that of the gluon self-energy contributions. 
By lengthy but straightforward calculation, 
it can be explicitly checked 
that they exactly cancel the $n$-dependent part of $i{\cal M}^{(b)}$ 
and the difference of the FP ghost loop contribution $i{\cal M}^{(c)}$ 
between the FD and Feynman gauges. 

Summing up, all the $n$-dependent terms in the scattering amplitudes for the process 
$q\bar{q}\to q^\prime\bar{q}^\prime$ cancel out exactly, and hence the FD gauge amplitudes 
agree exactly with those of the Feynman gauge in the one-loop order. 
\section{Use of background-field gauge fixing}
\label{sec:bkg}

In the preceding section, we have seen that loop integrals in the FD gauge have 
UV divergent parts including terms with an unconventional $i/\pi$ factor. 
These terms 
eventually cancel out in the total amplitudes. 
Moreover, the calculation of the UV-finite parts is even more difficult. 
These observations 
suggest that the FD gauge might be, although very useful at the tree level,
not suitable for loop calculation. 

Here we introduce an alternative method to include loop correction to the 
FD gauge amplitudes: the background-field gauge fixing 
method \cite{Kluberg-Stern:1974nmx,Kluberg-Stern:1975ebk,Abbott:1980hw,
Abbott:1981ke}, 
which may avoid the difficulties of the FD gauge loops while keeping its advantages 
at the tree level, as explained below. 

In the background-field gauge, 
the gluon field $A^a_{\mu}$ is expressed as a sum of the 
classical field $\tilde{A}^a_{\mu}$ and 
the quantum field $\hat{A}^a_{\mu}$ as 
$A^a_{\mu}\to\tilde{A}^a_{\mu}+\hat{A}^a_{\mu}$, and perform path integrals over 
quantum $\hat{A}^a_{\mu}$ around the background $\tilde{A}^a_{\mu}$. 
The effective action $\tilde{\Gamma}[\tilde{A}]\equiv\Gamma[\hat{A}=0, \tilde{A}]$ 
is then calculated from 1PI diagrams 
where all internal propagators are those of quantum fields, 
while all external fields are classical ones. 

In the calculation of $\Gamma[\tilde{A}]$, 
we need to fix the gauge only for quantum gauge fields. 
On the other hand, the gauge fixing for $\tilde{A}$ is only necessary to 
give the propagator for $\tilde{A}$ in constructing 
scattering amplitudes from the effective action. 
Therefore, no theoretical problem arises by adopting different 
gauge fixing methods for classical and quantum gauge fields. 

The background-field gauge method adopts the following 
function to fix the gauge for the quantum field $\hat{A}$ 
\begin{equation}
\tilde{F}^a[\hat{A}, \tilde{A}] = (\tilde{D}^{\mu}\hat{A}_{\mu})^a 
= \partial^{\mu}\hat{A}^a_{\mu} -g f^{abc}\tilde{A}^{b\mu}\hat{A}^c_{\mu} .
\label{bfm1} 
\end{equation}
with the gauge fixing term  
\begin{equation}
{\cal L}_{GF,BFG}[\hat{A}, \tilde{A}]
= -\frac{1}{2\xi_Q} (\tilde{F}^a[\hat{A}, \tilde{A}] )^2 ,
\label{bfm1b} 
\end{equation}
and the corresponding FP ghost Lagrangian 
\begin{equation}
{\cal L}_{FP,BFG}[\hat{A}, \tilde{A}] = 
i\bar{c}^a\tilde{D}^{\mu} (D_{\mu} c)^a .
\label{bfm1c}
\end{equation}
This gauge fixing preserves invariance under the ``classical''
gauge transformation, 
\begin{equation}
\tilde{\delta} \hat{A}^a_{\mu} = -g f^{abc}\omega^b \hat{A}^c_{\mu}, \; 
\tilde{\delta}\tilde{A}^a_{\mu} = -g f^{abc}\omega^b \tilde{A}^c_{\mu} 
-\partial_{\mu} \omega^a , 
\label{bfm2}
\end{equation}
where $\omega^a(x)$ are infinitesimal phases, 
but breaks invariance under the ``quantum'' gauge transformation, 
\begin{equation}
\hat{\delta} \hat{A}^a_{\mu} =  -g f^{abc}\omega^b (\tilde{A}^c_{\mu}+\hat{A}^c_{\mu})  
-\partial_{\mu} \omega^a, \; 
\hat{\delta}\tilde{A}^a_{\mu} = 0.
\label{bfm3} 
\end{equation}
As a result, the effective action $\tilde{\Gamma}[\tilde{A}]$ is  
manifestly invariant under the classical gauge transformation (\ref{bfm2}).  
In particular, the gluon self energy $\tilde{\Pi}_{\mu\nu}(q)$ and 
$q\bar{q}g$ vertex function $\tilde{\Gamma}^{\mu}(q,p_1,p_2)$ for on-shell quarks 
satisfy $q^{\mu}\tilde{\Pi}_{\mu\nu}(q)=0$ and 
$q_{\mu}\tilde{\Gamma}^{\mu}(q,p_1,p_2)=0$ for general $q$. 
Furthermore, since 
calculation of $\tilde{\Gamma}[\tilde{A}]$ is manifestly 
Lorentz covariant, for an arbitrary $\xi_Q$, 
we may express the self energy as 
\begin{equation}
\tilde{\Pi}_{\mu\nu}(q) = \left( -g_{\mu\nu} + \frac{ q_{\mu}q_{\nu} }{q^2} \right) 
\tilde{\Pi}_T(q^2)  .
\label{bfm4}
\end{equation}
It is then clear that, if $\tilde{\Pi}_{\mu\nu}$ and $\tilde{\Gamma}^{\mu}$ are 
used in place of $\Pi_{\mu\nu}$ and $\Gamma^{\mu}$, 
eqs.~(\ref{ampbc}, \ref{mvertex3}) do not depend on whether the Feynman gauge or 
FD gauge is used for the propagator $iP_{\mu\nu}(q)/q^2$ 
connecting the 1PI amplitudes. 

The one-loop gluon contributions to the gluon self energy $\tilde{\Pi}_{\mu\nu}$ and the
$q{q}g$ vertex function $\tilde{\Gamma}^{\mu}$ in the background-field gauge 
are given in 
refs.\,\cite{Kluberg-Stern:1974nmx,Kluberg-Stern:1975ebk,Abbott:1980hw}. 
Their UV divergences are 
\begin{align}
i\tilde{\Pi}_{\mu\nu}^{(b)}(q)|_{div}  &= i\frac{C_A g^2}{(4\pi)^2\epsilon}
\left( -q^2g_{\mu\nu} + q_{\mu}q_{\nu} \right) \left( -\frac{10}{3} \right) ,  
\label{bfmself1}  
\\
i\tilde{\Pi}_{\mu\nu}^{(c)} (q)|_{div} 
&= i\frac{C_A g^2}{(4\pi)^2\epsilon} 
\left( -q^2g_{\mu\nu} + q_{\mu}q_{\nu} \right) \left( -\frac{1}{3} \right) ,  
\label{bfmself2}  
\end{align}
and
\begin{align}
i\tilde{\Gamma}^{(d) \mu}|_{div} &= 
i \frac{C_A g^3}{(4\pi)^2\epsilon} \gamma^{\mu} \left( -\frac{\xi_Q}{2} \right) , 
\\
i\tilde{\Gamma}^{(e+f) \mu}|_{div} &= 
i \frac{C_A g^3}{(4\pi)^2\epsilon} \gamma^{\mu} \left( +\frac{\xi_Q}{2} \right) , 
\end{align}
respectively. Note that the UV divergence 
(\ref{bfmself1}) is independent of the gauge parameter $\xi_Q$. 
We also note that the total $q{q}g$ vertex function is UV finite and 
that the renormalization of the gauge coupling is entirely given by the 
gluon self energy \cite{Abbott:1980hw}. 

We finally comment on the resummation of the gluon self energy contribution. 
In the case where 
the gluon self energy takes the form (\ref{bfm4}), we may resum 
its contributions to the gluon propagator in the FD gauge by the replacement 
\begin{equation}
i\frac{P^{\rm FD}_{\mu\nu}(q)}{q^2} \to i\frac{P^{\rm FD}_{\mu\nu}(q)}{q^2+\tilde{\Pi}_T(q^2)} .
\label{resum1}
\end{equation}
This is proved by using the relation 
\begin{equation}
i\frac{P^{\rm FD}_{\mu\rho}(q)}{q^2}  \left( -g^{\rho\sigma} 
+ \frac{ q^{\rho}q^{\sigma} }{q^2} \right) 
i\tilde{\Pi}_T(q^2) 
\,
 i\frac{P^{\rm FD}_{\sigma\nu}(q)}{q^2}
=
- i\frac{P^{\rm FD}_{\mu\nu}(q)}{q^4} \tilde{\Pi}_T(q^2) . 
\label{resum2}
\end{equation}
Because the self-energy correction $\tilde{\Pi}_T(q^2)$ is 
the only UV divergent 1PI amplitudes at one-loop, 
giving the beta function of $g$, the identity (\ref{resum1}) 
may pave the way to improve the tree-level 
amplitudes in the FD gauge, given e.g. in Ref.~\cite{Hagiwara:2020tbx}, 
simply by replacing the gauge couplings 
by the  running couplings. 

\section{Summary}\label{sec:summary}
We have studied radiative corrections in the Feynman-Diagram (FD) 
gauge \cite{Hagiwara:2020tbx,Chen:2022gxv,Chen:2022xlg},
where the gauge boson is quantized along the light cone facing
the opposite of its three momentum, eq.~(\ref{FDlightcone}).
We have calculated the QCD scattering amplitudes for the process
$q\bar{q}\to q'\bar{q}'$ at one-loop level, and obtained the following
results:
\begin{itemize}
\item
The FP ghosts do not decouple from the scattering amplitudes
because the light-cone vector in the FD gauge depends on the
three momentum of gluons.
\item
Loop integrals cannot be done by conventional methods because
of the non-analyticity of the integrand. 
\item
UV singularities with a factor of $i/\pi$ times the conventional
ones appear from the $1/(n(k)\cdot k)$ factor, which does not
have a pole in the FD gauge. 
\item 
When the FD gauge propagators are expressed 
as the sum of the transverse (T) and 
the longitudinal (L) components, 
all the non-conventional UV singularities appear 
in the LL combinations of the two virtual gluons in the 
$q\bar{q}$ rest frame.
\item
All the non-conventional UV singularities cancel in the scattering
amplitudes when we sum over terms in the gluon and ghost loop
contribution to the propagator corrections, as well as those in the
initial $qqg$ and the final $q'q'g$ vertex corrections, reproducing
the known QCD beta function. 
\item
We have shown that the finite part of the radiative corrections is
identical to that of the Feynman gauge, because all the terms
which depend on the light-cone vector $n^{\mu}(q)$ cancel out
among 2-, 3-, and 4-point corrections. 
\end{itemize}

Summing up, we have reproduced the known QCD scattering amplitudes
for the process $q\bar{q} \to q^\prime\bar{q}^\prime$ at the one-loop level
in the FD gauge.
This has been proven by showing cancellation of all UV singularities
and the finite correction terms which depend on the light-cone
vector $n^{\mu}(q)$.

Although our findings suggest that the FD gauge is a consistent
gauge fixing for quantizing gluons, the lack of covariance and analyticity in
the regularized loop integrals does not allow us to take advantage
of the standard loop integral tools.
Instead, we propose that all the 1PI loop integrals should be
done in the Feynman gauge on the FD gauge gluon background. 
We obtain the same 2- and 3-point loop functions as those of
the conventional background-field gauge, 
in which both the quantum and background 
gluons are in the Feynman gauge.
Schwinger-Dyson summation of all the one-loop propagator
corrections connected by FD gauge gluons gives the one-loop
corrected FD gauge propagator. 
The results may be useful in obtaining improved Born 
approximation to the tree-level FD gauge amplitudes.
\section*{Acknowledgements}
  We thank Daniel Chung for illuminating discussions.
  KH and YJZ wish to thank Vernon Barger for discussions
  and hospitality at University of Wisconsin-Madison,
  where part of the work has been done.
All the Feynman diagrams were drawn with 
{\tt TikZ-FeynHand}~\cite{Ellis:2016jkw,Dohse:2018vqo}.
The work was supported in part by, JSPS KAKENHI Grant No.\,21H01077, 21K03585, 23K03403, 
24K07032 and US Japan Cooperation Program in High Energy Physics.

\appendix
\renewcommand{\thesection}{\Alph{section}}
\section{Calculation of loop integrals in the FD gauge}
\label{sec:AppA}

In this appendix, we explain how we 
evaluate the UV singular parts of loop integrals 
with factors $n(k)\cdot k=|k^0|+|\vec{k}\,|$ in the denominator. 

We first note that the UV divergences of loop integrals in our self energy and 
vertex corrections should be polynomials of external momenta, to appropriate order. 
For a gluon self energy loops in $\Pi_{\mu\nu}(q)$, for example, we 
differentiate the integrands two times 
by $q^{\mu}=(q^0,q^i)$ and 
take $q\to 0$, after regularizing the integrands to avoid 
infrared divergences generated by these operations. 
We may then perform loop integrations, which are not Lorentz covariant in general, 
by known techniques. 
The UV-divergent parts of the original loops are then easily obtained. 

For illustration, we calculate the UV divergence of the 0 component of 
the integral eq.~(\ref{sampleintegral}),  
\begin{equation}
I^0(q)= \int \frac{d^D k}{(2\pi)^D} \frac{ {\rm sgn}(k^0) }{ 
(|k^0|+|\vec{k}\,|) ((k+q)^2 + i0) } .
\label{sampleintegral2}
\end{equation}
By dimension counting and the $(D-1)$ space dimensional rotational invariance, we can tell that 
its UV divergent part should take the form 
$a_0 q^0$ with a $q$-independent coefficient $a_0$. 

We first differentiate $I^0(q)$ by $q^0$ to obtain 
\begin{equation}
\frac{\partial I^0}{\partial q^0} (q) = \int \frac{d^D k}{(2\pi)^D} 
\frac{ -2(k^0+q^0) {\rm sgn}(k^0)  }
{  (|k^0|+|\vec{k}\,|) ((k+q)^2 + i0)^2 } .
\label{sampleintegra3}
\end{equation}
By using the factorization
\begin{align}  
(k+q)^2 + i0 & =(k^0+q^0)^2-|\vec{k}+\vec{q}\,|^2 +i0
\nonumber \\
& =(|k^0+q^0|+|\vec{k}+\vec{q}\,|)(|k^0+q^0|-|\vec{k}+\vec{q}\,|+i0 ) ,
\label{factorization}
\end{align}
the denominators of the integrands become products of $(|l^0|+|\vec{l}\,|)$ 
and $(|l^0|-|\vec{l}\,|+i0)$ 
($l$: a momentum of the propagator). 
After introducing a fictitious mass parameter $m>0$ as 
$(|l^0|\pm|\vec{l}\,|)\to(|l^0|\pm(|\vec{l}\,|+m))$ 
to avoid infrared divergences,  
we take $q\to0$ limit to obtain 
\begin{equation}
\frac{\partial I^0}{\partial q^0} (0) = \int \frac{d^D k}{(2\pi)^D} 
\frac{ -2 |k^0|  }
{  (|k^0|+|\vec{k}\,|+m)^3 ( |k^0| -|\vec{k}\,|-m + i0)^2 } .
\label{sampleintegral4}
\end{equation}
We then perform $(D-1)$-dimensional space integration 
by using 
\begin{eqnarray}
\frac{d^{D-1}\vec{k}}{(2\pi)^{D-1}} \to \frac{1}{(4\pi)^{\frac{D-1}{2}}} 
\frac{2}{\Gamma\left(\frac{D-1}{2}\right) } |\vec{k}\,|^{D-2} d|\vec{k}\,| ,
\label{eq:D-1int}
\end{eqnarray}
after decomposing eq.~(\ref{sampleintegral4}) into  sum of 
fractions $1/(|\vec{k}\,|+A)^n$, where $A=|k^0|+m$ or $-|k^0|+m-i0$. 
For example, the integration of 
$1/(|\vec{k}\,|+A)$ is 
\begin{align}
\int \frac{d^{D-1}\vec{k}}{(2\pi)^{D-1}} \frac{1}{|\vec{k}\,|+A} & = 
\frac{1}{(4\pi)^{\frac{D-1}{2}}}
\frac{2}{\Gamma\left(\frac{D-1}{2}\right) } \frac{-\pi}{\sin(D\pi)} 
A^{D-2} 
\nonumber \\
& = \frac{1}{(4\pi)^2} \left( \frac{4}{\epsilon} + O(\epsilon^0) \right)  A^{D-2},
\label{eq:D-1int2}
\end{align}
where $D=4-2\epsilon$. Integration of  $1/(|\vec{k}\,|+A)^n$ for $n\ge 2$ is then obtained by 
differentiating eq.~(\ref{eq:D-1int2}) by $A$. 

The space integration in \eqref{sampleintegral4} then takes the form 
\begin{equation}
\frac{1}{(4\pi)^2}\left( \frac{4}{\epsilon}+O(\epsilon^0) \right) 
\int_{0}^{\infty} \frac{d |k^0|}{\pi} 
\left[ f_1(|k^0|, m)(m+|k^0|)^{D-4} + f_2(|k^0|, m)(m-|k^0|-i0)^{D-4} \right] , 
\label{sampleintegral5}
\end{equation}
where $f_{1,2}$ are rational functions of $|k^0|$ and $m$:
\begin{eqnarray}
f_1(|k^0|, m)&=& 
\frac{1}{8|k^0|^3} 
\left[ (-2D^2+14D-23) |k^0|^2 
+ 2(2D-7)m|k^0| -3m^2 \right] ,
\label{samplef1}
\\
f_2(|k^0|, m)&=&
\frac{1}{8|k^0|^3}
\left[ (-2D+7)|k^0|^2 +2(D-5)m|k^0| 
+3m^2 \right] ,
\label{samplef2}
\end{eqnarray}
for the integral\,\eqref{sampleintegral4}.
Note that the term of order $|k^0|^{-1}$ 
in eqs.~(\ref{samplef1}) and (\ref{samplef2}) cancel 
in eq. (\ref{sampleintegral5}) in the $D=4$ limit, 
and hence the $|k^0|$ integral in (\ref{sampleintegral5}) 
is UV finite. 
The factor $(m-|k^0|-i0)^{D-4}$ for  $|k^0|>m$ should be interpreted as 
$(|k^0|-m)^{D-4}\exp(-i(D-4)\pi)$.  
The $|k^0|$ integration in eq.~(\ref{sampleintegral5}) can then be performed 
analytically by 
splitting the integration region into $(0, m)$ and $(m, \infty)$, 
remembering that $D$ is a general complex number. 

The integration of \eqref{sampleintegral5} 
in $0\le |k^0| \le m$ is, using $|k^0|=mx$, 
\begin{equation}
\frac{1}{(4\pi)^2}\left( \frac{4}{\epsilon}+O(\epsilon^0) \right) \frac{1}{\pi} m^{D-4} 
\int_{0}^{1} dx \, 
\left[ f_1(x, 1)(1+x)^{D-4} + f_2(x, 1)(1-x)^{D-4} \right]  . 
\label{sampleintegral5a}
\end{equation}
It is seen that there is no singularity 
in \eqref{sampleintegral5a},  
including the boundaries at $x=0$ and $x=1$, 
for $D\simeq 4$.  
On the other hand, substitution of $D=4$ into the integrand of 
(\ref{sampleintegral5a}) just gives 0. So, the integral should be $O(\epsilon)$, 
irrelevant in our calculation.  

The integration for the other part,  $m\le |k^0|<\infty$, is written as 
\begin{equation}
\frac{1}{(4\pi)^2}\left( \frac{4}{\epsilon}+O(\epsilon^0) \right) \frac{1}{\pi} m^{D-4} 
\int_{1}^{\infty} dx \, 
\left[ f_1(x, 1)(x+1)^{D-4} + f_2(x, 1)(x-1)^{D-4} e^{-i(D-4) \pi} \right] .
\label{sampleintegral5b}
\end{equation}
Since $f_{1,2}(x,1)\to O(1/x)$ for $x\to\infty$, the integral (\ref{sampleintegral5b}) is 
apparently divergent for $D\simeq 4$, but again the integrand vanishes at $D=4$. 
We therefore expect that integral of (\ref{sampleintegral5b})  gives 
a finite result as $(D-4)\times\frac{1}{(D-4)}$. 

Let us calculate $(x+1)^{D-4}$ part of the integral in  (\ref{sampleintegral5b}) , 
\begin{equation}
I_1 = \int_1^{\infty} dx \, f_1(x,1)(x+1)^{D-4} , 
\end{equation}
first. 
By decomposing $f_1(x,1)$ as 
$C/(x+1)+O(1/(x+1)^2)$, where $C$ is a function of $D$, the integrand is 
written as 
\begin{equation}
\frac{-2D^2+14D-23}{8}(x+1)^{D-5} + \frac{(-2D^2+18D-37)x^2+ (4D-17)x-3}{8x^3}
(x+1)^{D-5} . 
\label{sampleintegral5b1}
\end{equation}
The first term gives a divergence. By using 
\begin{equation}
\int_1^{\infty} dx \, (x+1)^{D-5} = -\frac{2^{D-4}}{D-4} ,
\end{equation}
it is
\begin{equation}
-\frac{1}{8(D-4)} +\frac{1}{4}-\frac{1}{8}\log 2 + O(D-4) .
\label{sampleintegral5b2}
\end{equation}
The second term behaves as $1/x^2$. One can therefore evaluate its finite term 
by substituting $D=4$.  The result is 
\begin{equation}
\frac{1}{16}+\frac{1}{8}\log 2 + O(D-4) . 
\label{sampleintegral5b3}
\end{equation}
By adding (\ref{sampleintegral5b2}) and (\ref{sampleintegral5b3}), we obtain 
\begin{equation}
I_1 = -\frac{1}{8(D-4)}+ \frac{5}{16} + O(D-4) .
\label{sampleintegral5c1}
\end{equation}

Next, we calculate $(x-1)^{D-4}$ part of the integral in  (\ref{sampleintegral5b}) ,
\begin{equation}
I_2 = \int_1^{\infty} dx \, f_2(x,1)(x-1)^{D-4} , 
\end{equation}
where the factor $e^{-i(D-4)\pi}$ will be put in later. 
Although we are going to remove $O(1/(x-1))$ part from 
$f_2(x,1)$ as was done for $f_1$, we have to avoid generating a singularity at $x=1$ 
by this subtraction. For this purpose, we split the integration region into $1\le x \le 2$ and 
$2\le x<\infty$. The former integral gives 
\begin{equation}
\frac{1}{64} - \frac{1}{8}\log 2 +O(D-4). 
\label{sampleintegral5b5} 
\end{equation}
For  the latter integral, we split the $C/(x-1)$ part from $f_2(x,1)$. 
The integrand is then written as 
\begin{equation}
\frac{-2D+7}{8}(x-1)^{D-5} + \frac{(4D-17)x^2+(-2D+13)x-3}{8x^3}(x-1)^{D-5} .
\end{equation}
The first term gives 
\begin{equation}
\frac{1}{8(D-4)}+ \frac{1}{4} + O(D-4) , 
\label{sampleintegral5b6}
\end{equation}
by using
\begin{equation}
\int_2^{\infty} dx \, (x-1)^{D-5} = -\frac{1}{D-4} .
\end{equation}
The second term can be calculated in $D\to 4$, giving 
\begin{equation}
-\frac{5}{64}+\frac{1}{8}\log 2 +O(D-4) . 
\label{sampleintegral5b7}
\end{equation}
Summation of eqs. (\ref{sampleintegral5b5}, \ref{sampleintegral5b6}, 
\ref{sampleintegral5b7}) gives 
\begin{equation}
I_2 = \frac{1}{8(D-4)} +\frac{3}{16} + O(D-4) . 
\label{sampleintegral5c2}
\end{equation}
By inserting (\ref{sampleintegral5c1}) and 
(\ref{sampleintegral5c2}) in the integral 
(\ref{sampleintegral5}), we obtain 
\begin{align}
\frac{\partial I^0}{\partial q^0}(0) & = 
\frac{1}{(4\pi)^2}\left( \frac{4}{\epsilon}+O(\epsilon^0) \right) \frac{1}{\pi} m^{-2\epsilon} 
( I_1 + I_2 e^{-i(D-4)\pi}  ) 
\nonumber \\ 
&= 
\frac{1}{(4\pi)^2}\left( \frac{4}{\epsilon}+O(\epsilon^0) \right) \frac{1}{\pi} m^{-2\epsilon} 
\left( \frac{1}{2} - \frac{i\pi}{8}+ O(\epsilon)  \right) 
\nonumber \\
&= \frac{1}{(4\pi)^2\epsilon}\left( \frac{2}{\pi}- \frac{i}{2} \right)+O(\epsilon^0) .
\label{sampleintegral6}
\end{align}
The integral $I^0(q)$ of eq.~(\ref{sampleintegral2}) is hence 
\begin{equation}
 I^0(q) = \frac{1}{(4\pi)^2\epsilon}
 \left( \frac{2}{\pi}- \frac{i}{2} 
 \right) q^0 +O(\epsilon^0) .
\end{equation}

All of the UV singular parts of the loop integrals 
involving $n^{\mu}$, which appear in Section 
\ref{sec:amplitude}, can be evaluated 
in the same manner. As another example, 
we calculate the UV singular part of 
the space components of the 
integral \eqref{sampleintegral}, 
\begin{equation}
I^j(q)= \int \frac{d^D k}{(2\pi)^D} \frac{ 1 }{ 
(|k^0|+|\vec{k}\,|) (k+q)^2 }
\frac{-k^j}{|\vec{k}\,|}  . 
\label{sampleintegralj}
\end{equation}
Its UV-singular part 
should take the form $a_1q^j$ with a $q$-independent coefficient $a_1$. 
In this case, we differentiate $I^j(q)$ by $q^i$, 
\begin{equation}
\frac{\partial I^j}{\partial q^i} (q) = \int \frac{d^D k}{(2\pi)^D} 
\frac{ -2(k^i+q^i) k^j  }
{  (|k^0|+|\vec{k}\,|) |\vec{k}\,| ((k+q)^2 + i0)^2 } .
\label{sampleintegraj3}
\end{equation}
Again, by introducing an IR regulator mass $m$ and 
factorization (\ref{factorization}), and 
taking the limit $q\to 0$,  we obtain 
\begin{align}
\frac{\partial I^j}{\partial q^i} (0) & = \int \frac{d^D k}{(2\pi)^D} 
\frac{ -2 k^i k^j  }
{  (|k^0|+|\vec{k}\,|+m)^3 ( |k^0| -|\vec{k}\,|-m + i0)^2 |\vec{k}\,| }
\nonumber \\
& =  \frac{1}{D-1} \int \frac{d^D k}{(2\pi)^D} 
\frac{ -2 |\vec{k}\,| \delta^{ij}  }
{  (|k^0|+|\vec{k}\,|+m)^3 ( |k^0| -|\vec{k}\,|-m + i0)^2 } .
\label{sampleintegralj4}
\end{align}
In the second line of eq. (\ref{sampleintegralj4}) we replaced 
$k^ik^j$ by $|\vec{k}\,|^2 \delta^{ij}/(D-1)$ by using the $(D-1)$ dimensional rotational invariance. 

The UV singular part of eq. (\ref{sampleintegralj4}) can be calculated 
in the same manner as that of $I^0$. 
The final result is 
\begin{equation}
I^j(q) = \frac{1}{(4\pi)^2\epsilon}\left( -\frac{2}{3\pi}- \frac{i}{2} \right) q^j 
+O(\epsilon^0) .
\end{equation}

\section{Transverse and longitudinal contributions to $\Pi_{\mu\nu}(q)$ 
for general $q^{\mu}$ }
\label{sec:AppB}

The UV divergence of the 
gluon self energy $i\Pi_{\mu\nu}(q)$ for general $q^{\mu}$ is separated 
into TT, TL, and LL parts as 
\begin{align}
i\Pi^{(bTT)}_{[00,0j,ij]}(q)|_{div} & 
= C_A g^2 \frac{1}{(4\pi)^2 \epsilon}  \left[ 
\frac{5i}{3}|\vec{q}\,|^2, \; 
-\frac{i}{3}q^0 {q}^j, \; 
\frac{i}{3} (q^0)^2 \delta^{ij} +\frac{9i}{5}|\vec{q}\,|^2 \delta^{ij}  
-\frac{31i}{15} {q}^i {q}^j \right] ,
\label{eq16a}
\\
i\Pi^{(bTL)}_{[00,0j,ij]}(q)|_{div} & 
= C_A g^2 \frac{1}{(4\pi)^2\epsilon} \left[ 
-\frac{16i}{3} |\vec{q}\,|^2, \;  
\left( \frac{10i}{3} + \frac{4}{3\pi} \right) q^0 {q}^j, \; 
\right. \nonumber \\
& \left. 
-\frac{8i}{3} (q^0)^2 \delta^{ij} +\left( \frac{8i}{15}-\frac{16}{15\pi} \right) 
|\vec{q}\,|^2 \delta^{ij}
+ \left( -\frac{4i}{15}+\frac{8}{15\pi} \right) {q}^i {q}^j
 \right] ,  \label{eq21a}
\\
i\Pi^{(bLL)}_{[00,0j,ij]}(q)|_{div} & 
= C_A g^2 \frac{1}{(4\pi)^2\epsilon} \left[ 
\frac{10}{3\pi}(q^0)^2-\frac{74}{9\pi}|\vec{q}\,|^2 , \; 
\frac{8}{9\pi} q^0 {q}^j, \; 
 \frac{22}{9\pi}(q^0)^2 \delta^{ij} -\frac{86}{45\pi} |\vec{q}\,|^2 \delta^{ij}
+\frac{68}{45\pi} {q}^i {q}^j  \right] .
\label{eq25a}
\end{align}
In contrast to the $q^{\mu}=(Q,\vec{0}\,)$ case (\ref{eq21}), 
the TL contribution (\ref{eq21a}) has both $O(i)$ and $O(1/\pi)$ terms.


%

\bibliography{bibfdloop}
\bibliographystyle{utphys}

\end{document}